\documentclass[aip,cha,showkeys,longbibliography,reprint,twocolumn,nofootinbib,floatfix] 
{revtex4-2}

\usepackage[utf8]{inputenc}
\usepackage[OT4]{fontenc}
\usepackage{graphicx}
\usepackage{amsmath}
\usepackage{dcolumn}
\usepackage{multirow}
\usepackage{hyperref}
\hypersetup{colorlinks=true}
\usepackage{cleveref}

\begin{document}
\title{Recovering Zipf's law in intercontinental scientific cooperation}  
\author{Malgorzata J. Krawczyk}
\email{malgorzata.krawczyk@agh.edu.pl}
\author{Krzysztof Malarz}
\thanks{Corresponding author}
\email{malarz@agh.edu.pl}
\affiliation{AGH University, Faculty of Physics and Applied Computer Science,
al. Mickiewicza 30, 30-059 Krak\'ow, Poland}
\date{July 6, 2023}%
\revised{September 23, 2023}%

\begin{abstract}
Scientific cooperation on an international level has been well studied in the literature. However, much less is known about this cooperation on the intercontinental level. In this paper, we address this issue by creating a collection of approximately $13.8$ million publications around the papers by one of the highly cited author working in complex networks and their applications. The obtained rank-frequency distribution of the probability of sequences describing continents and number of countries---with which authors of papers are affiliated---follows the power law with an exponent $-1.9108(15)$. Such a dependence is known in the literature as Zipf's law and it has been originally observed in linguistics, later it turned out that it is very commonly observed in various fields.
The number of distinct ``continent (number of countries)'' sequences in a function of the number of analyzed papers grows according to power law with exponent $0.527(14)$, i.e. it follows Heap's law.
\end{abstract}

\keywords{scientific papers and databases; scientific cooperation; Zipf's law; Heap's law}

\maketitle

\begin{quotation}
Collaboration between scientists seems to be an important issue in the scientific world. 
However, the question is what kind of cooperation are we talking about. 
Does it only concern colleagues from the same institution, country, or is it wider and includes people from very different research centers? 
To answer this question, we have selected a collection of nearly 14 million publications and are analyzing it in terms of international and intercontinental scientific cooperation.

Our analysis is based on mapping the authors to the continents, where the institutions indicated in the authors' affiliations of particular articles are located. 
This allows us to obtain ranks reflecting the frequency distribution of the probability of occurrence of specific alphabetically ordered sequences ``continent (number of countries)''.
The sequences describe the continents (including the number of countries in this continent) from which scientists publishing together come from.

The most frequently observed sequences [``North America (1)'', ``Europe (1)'', ``Asia (1)'' and ``Europe (2)''] obtained from mapping authors’ countries to continents do not show their intercontinental nature.
The first intercontinental cooperation [``Europe (1), North America (1)''] is ranked as the fifth and corresponds to papers written by authors from a single European country plus authors from a single country in North America.
The next most frequent cases [``Asia (1), North America (1)'' and ``Asia (1), Europe(1)''] correspond to a bilateral intercontinental cooperation established by authors from a pair of countries.
At seventh rank we meet again papers from single country from Australia \& Oceania.

The probability distributions of occurring sequences ``continent (number of countries)'' based on the obtained rank frequency reveals the dependence of the power law with an exponent $\mathbf{-1.9108(15)}$, known as Zipf's law, and occurs interdisciplinarily in many types of studied data ranging from quantitative linguistics (where it was originally formulated), via social sciences, music, seismic activity, physics, and others.
The number of various ``continent (number of countries)'' sequences in a function of the number of analyzed papers grows according to power law with exponent $\mathbf{0.527(14)}$, i.e. it follows Heap's law. 

\end{quotation}

\section{Introduction}

Studies of scientific cooperation between authors of scientific articles are mainly focusing on the international level \cite{WOS:000084641800003,*WOS:000233960500009,*WOS:000260903700007,*WOS:A1992GW95200006,*WOS:A1996VG50500006,*WOS:A1997XC57700001,*WOS:000302157900009,*WOS:000169909700005,*WOS:000249187700005,*WOS:A1993LW52100002}.
Much less is known about intercontinental scientific cooperation.
In the Web of Science database \cite[][accessed on March 2023]{WOS} we were able to identify only two articles \cite{Kozma_2019,Abramo_2020} dealing with this topic.

In Ref.~\onlinecite{Kozma_2019} the analysis focuses on mapping, based on publication output, scientific cooperation of African researchers, and the role of the South African research community as a channel for within- and intercontinental collaborations.
\citeauthor*{Kozma_2019} selected ten scientific fields (tropical medicine; parasitology; infectious diseases; ecology; water resources; immunology; zoology; plant sciences; agricultural and food sciences; and psychology) to illustrate and analyze scientific activity.
The authors created cooperation networks and visualized them on world maps.
In addition, they calculated measures of network centrality to test the frequency of involvement of different countries, with particular emphasis on South Africa, in the collaborative process.
In addition, authors appearing in the first or last position on the list of authors of publications were thrown into one bag in order to examine the influence of selected authors on research directions and their contribution to publications. Finally, the most important research funding agencies and their geographical distribution were distinguished. By combining these stages of analysis, the authors obtained a precise picture of the level of involvement of the South African research community in both intra- and intercontinental scientific cooperation.

In Ref.~\onlinecite{Abramo_2020} the influence of geographic distance on knowledge flows, measured through citations to scientific publications was studied. \citeauthor*{Abramo_2020} showed that the influence of the physical distance between the authors of the papers (or rather the distance between the geographical coordinates of their institutions) on the flow of knowledge is significant at the national level, not insignificant at the continental level, but completely marginal at the intercontinental level.

To fill the gap mentioned above in continent-based studies of scientific cooperation, in this paper, we construct a set of nearly 13.8 million publications and analyze it in terms of intercontinental scientific cooperation.
Based on mapping authors' to continents (via the countries where their institutions are located), we generate sequences of continents with the number of countries on these continents.
As we show below, the obtained rank-frequency distribution of sequences reveals Zipf's law \cite{Zipf_1935} with an exponent close to $-1.9$.

\section{Methodology}

To collect data on the affiliation of authors of scientific publications, we used the Scopus API\footnote{application programming interface} [\onlinecite[][accessed on July 2022]{api}]. As we are interested in analysis of scientific cooperation, we decided to build a graph around scientist who is active in more than one discipline. 
The science that meets this condition is complex systems. To select the person who will serve us as a starting point for data collection, we used the list of the recipients of `Senior Scientific Award' of the Complex System Society in years 2014--2020 (see \Cref{tab:ssa_ccs}). 
 After calculating the average value of the number of citations per article and the value of the H-index, we selected the author \cite{WoS_Vespignani} for whom the values of these two parameters are closest to the average values.
His scientific interests cover the analysis of social \cite{WOS:000668492600013,*WOS:000567522200024,*WOS:000444788000001} and biological networks \cite{WOS:000229193300002,*WOS:000175859600017,*WOS:000220314500008,*WOS:000235464700021}, or the spread of epidemics \cite{WOS:000819659900005_at_all,*WOS:000805653200014,*WOS:000815971000001_et_all,*WOS:000742728600007,*WOS:000722150700001}.
Subsequently, we found all authors who published with him, then found their co-authors, and so on. However, during data aggregation, we decided to stop this recursive finding for authors whose total number of publications is less than fifty, for authors whose last publication was published before $2015$ and for authors whose distance, measured as the number of edges on a shortest path connecting a given author with the starting node, in the graph of authors is greater than six. 
From this set, we also excluded articles if the number of affiliations of a given author at a particular publication is greater than five or if the data associated with a given publication do not allow automated country identification.
This methodology allowed us to collect data on $13\,810\,394$ publications.

\begin{table*}[]
\caption{\label{tab:ssa_ccs}Recipients of `Senior Scientific Award' of the Complex System Society in years 2014--2020 \cite{ccs-soc}}
    \centering
\begin{ruledtabular}
\begin{tabular}{lcrrrr}
Year & Name & Publications & Times Cited & Average per item & H-Index \\\hline
2014 & H. Eugene Stanley\cite{WoS_Stanley}                & 1243 & 111503 &  89.70 & 154 \\
2015 & Maxi San Miguel\cite{WoS_Miguel}                  &  291 &   10651 &  36.60 &  50 \\
2016 & Michael Batty\cite{WoS_Batty}                    &  464 &  13017 &  28.05 &  63 \\
2017 & Albert-László Barabási\cite{WoS_Barabasi}           &  326 & 132200 & 403.05 & 133 \\
2018 & Alessandro Vespignani\cite{WoS_Vespignani}       &  274 &  44528 & 162.51 &  95 \\
2019 & Yamir Moreno\cite{WoS_Moreno}                    &  243 &  28255 & 116.27 &  67 \\
2020 & Jose Fernando Ferreira Mendes\cite{WoS_Mendes}   &  159 &  12117 &  76.21 &  40 \\
    \end{tabular}
\end{ruledtabular}
\end{table*}

Affiliations contained in each publication are processed as follows:
\begin{enumerate}
    \item for each author all his/her $n_i$ affiliations mentioned in a given paper are identified together with a country where this affiliation is settled;
    \item duplicates of countries in the list created for an individual author are removed, i.e., each country may appear in this list exactly once, which results in a number $n_c^a$ of countries per author;
    \item the lists of countries for all authors are merged into a single list---without repetitions---resulting in $n_c$ countries; 
    \item $n_c$ countries that appear in a given publication are then mapped to continents on which these countries are located;
    \item for the purposes of further analysis, a number is added to each continent in the collection, specifying the number of different countries located on this continent represented by the authors of a given publication.
\end{enumerate}

\begin{figure}[htbp]
\includegraphics[width=0.99\columnwidth]{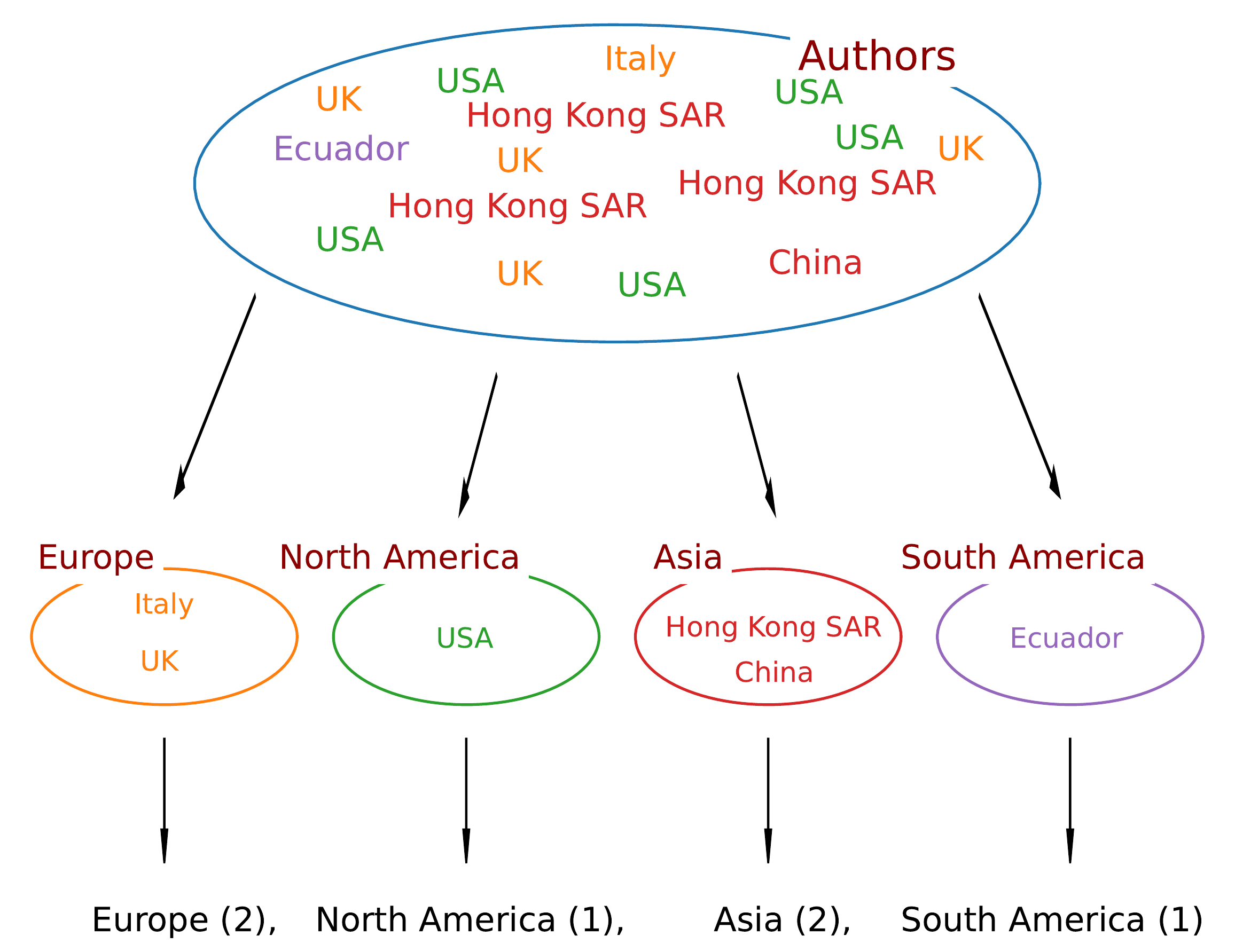}
\caption{\label{fig:mapping}Presentation of the method of mapping the affiliations of dozen authors (from six countries and four continents) of Ref.~\onlinecite{cc} to continents sequences. The sequence obtained is as follows: Asia (2), Europe (2), North America (1), South America (1)}
\end{figure}

For a better understanding of this process, let us analyze an example publication~\cite{cc}. In \Cref{tab:mapping_example} all their authors (in alphabetical order) are presented, with indication of the countries of particular affiliation, number of affiliations $n_i$ and number of countries $n_c^a$. Nine out of twelve authors of this work are affiliated with only one institution in one country. Two of the twelve authors (3rd and 11th) are affiliated with two institutions in two countries. The remaining one (4th) is affiliated with three institutions located in two countries. All these countries are in the oval (marked as ``Authors'') at the top of \Cref{fig:mapping}. In the next step, we map countries to continents (removing country repetitions) and sum up the number of countries on each continent.
Several examples of such mapping are presented in \Cref{tab:mapping}.

\begin{table}[htbp]
\caption{\label{tab:mapping_example}Presentation of the method of mapping $n_i$ affiliations of every author of Ref.~\onlinecite{cc} to $n^a_c$ countries}
\begin{ruledtabular}
\begin{tabular}{rlcc}
& author (country of affiliation) & $n_i$ & $n^a_c$\\\hline
1 & Christopher Dye (UK) & 1 & 1 \\
2 & Rosalind M. Eggo (UK) & 1 & 1 \\
3 & Bernardo Gutierrez (UK, Ecuador) & 2 & 2 \\
4 & Moritz U. G. Kraemer (UK, USA, USA) & 3 & 2 \\
5 & Gabriel M. Leung (Hong Kong SAR) & 1 & 1 \\
6 & Kathy Leung (Hong Kong SAR) & 1 & 1 \\
7 & James O. Lloyd-Smith (USA) & 1 & 1 \\
8 & Samuel V. Scarpino (USA) & 1 & 1 \\
9 & Munik Shrestha (USA) & 1 & 1 \\
10 & Huaiyu Tian (China) & 1 & 1 \\
11 & Alessandro Vespignani (Italy, USA) & 2 & 2 \\
12 & Joseph Wu (Hong Kong SAR) & 1 & 1 \\
\end{tabular}
\end{ruledtabular}
\end{table}

\begin{table}[htbp]
\caption{\label{tab:mapping}Some examples of mapping $n_a$ authors in $n_c$ countries to `continent (number of countries)' sequences. The examples are based on Refs. \onlinecite{cc,example_6_at_all,Krawczyk2005,Krawczyk2010,Stauffer2006,example_4,example_1,example_3,example_2,example_5,Lima2006}.
Also the sequence rank $\mathcal R$ according to Ref.~\onlinecite{data-intercont-rank} is presented}
\begin{ruledtabular}
\begin{tabular}{lrrp{55mm}r}
Ref.                & $n_a$ &  $n_c$ & sequence & $\mathcal R$ \\ 
\hline 
\onlinecite{cc}           & 12 & 6 & Asia (2), Europe (2), North America~(1), South America~(1) & 263\\
\hline 
\onlinecite{example_6_at_all}    & 49 & 12 & Africa (1), Asia (1), Europe (8), North America~(1), South America~(1) & 4374    \\
\hline 
\onlinecite{Krawczyk2005} &  5 & 1 & Europe (1)                    & 2 \\
\hline 
\onlinecite{Krawczyk2010} &  4 & 2 & Europe (2)                    & 4 \\
\onlinecite{Stauffer2006} &  3 & 2 & Europe (2)                    & 4 \\
\hline 
\onlinecite{example_4}    &  4 & 3 & Europe (3) & 11 \\
\hline 
\onlinecite{example_1}    &  4 & 2 & Europe (1), North America (1) & 5  \\
\hline 
\onlinecite{example_3}    &  4 & 3 & Europe (2), North America (1) & 10 \\
\onlinecite{example_2}    &  3 & 3 & Europe (2), North America (1) & 10 \\
\onlinecite{example_5}    &  3 & 3 & Europe (2), North America (1) & 10 \\
\hline 
\onlinecite{Lima2006}     &  2 & 2 & Europe (1), South America (1) & 19 \\
\end{tabular}
\end{ruledtabular}
\end{table}

\begin{table}[htbp]
\caption{\label{tab:zipf}Top-20 most frequent identified sequences. The complete list is available in Ref.~\onlinecite{data-intercont-rank}}
\begin{ruledtabular}
\begin{tabular}{r l d}
$\mathcal R$  & sequence & \%\\ \hline
1 & North America (1) & 26.51 \\
2 & Europe (1) & 24.59 \\
3 & Asia (1) & 17.06 \\
4 & Europe (2) & 4.95 \\
5 & Europe (1), North America (1) & 4.56 \\
6 & Asia (1), North America (1) & 3.66 \\
7 & Australia \& Oceania (1) & 2.08 \\
8 & Asia (1), Europe (1) & 1.99 \\
9 & Asia (2) & 1.26 \\
10 & Europe (2), North America (1) & 1.09 \\
\hline 
11 & Europe (3) & 1.02 \\
12 & North America (2) & 0.98 \\
13 & South America (1) & 0.77 \\
14 & Australia \& Oceania (1), Europe (1) & 0.58 \\
15 & Asia (1), Europe (1), North America (1) & 0.56 \\
16 & Asia (1), Australia \& Oceania (1) & 0.50 \\
17 & Australia \& Oceania (1), North America (1) & 0.47 \\
18 & Asia (1), Europe (2) & 0.39 \\
19 & Europe (1), South America (1) & 0.36 \\
20 & North America (1), South America (1) & 0.36
\end{tabular}
\end{ruledtabular}
\end{table}

\section{Results}

In \Cref{tab:zipf} the top-20 of the most frequent sequences in the form ``continent (number of countries)'' together with its percentage in all analyzed publications are presented.
The complete list (containing nearly nine thousand types of case) is available in Ref.~\onlinecite{data-intercont-rank}.

\begin{figure}[b]
\includegraphics[width=0.99\columnwidth]{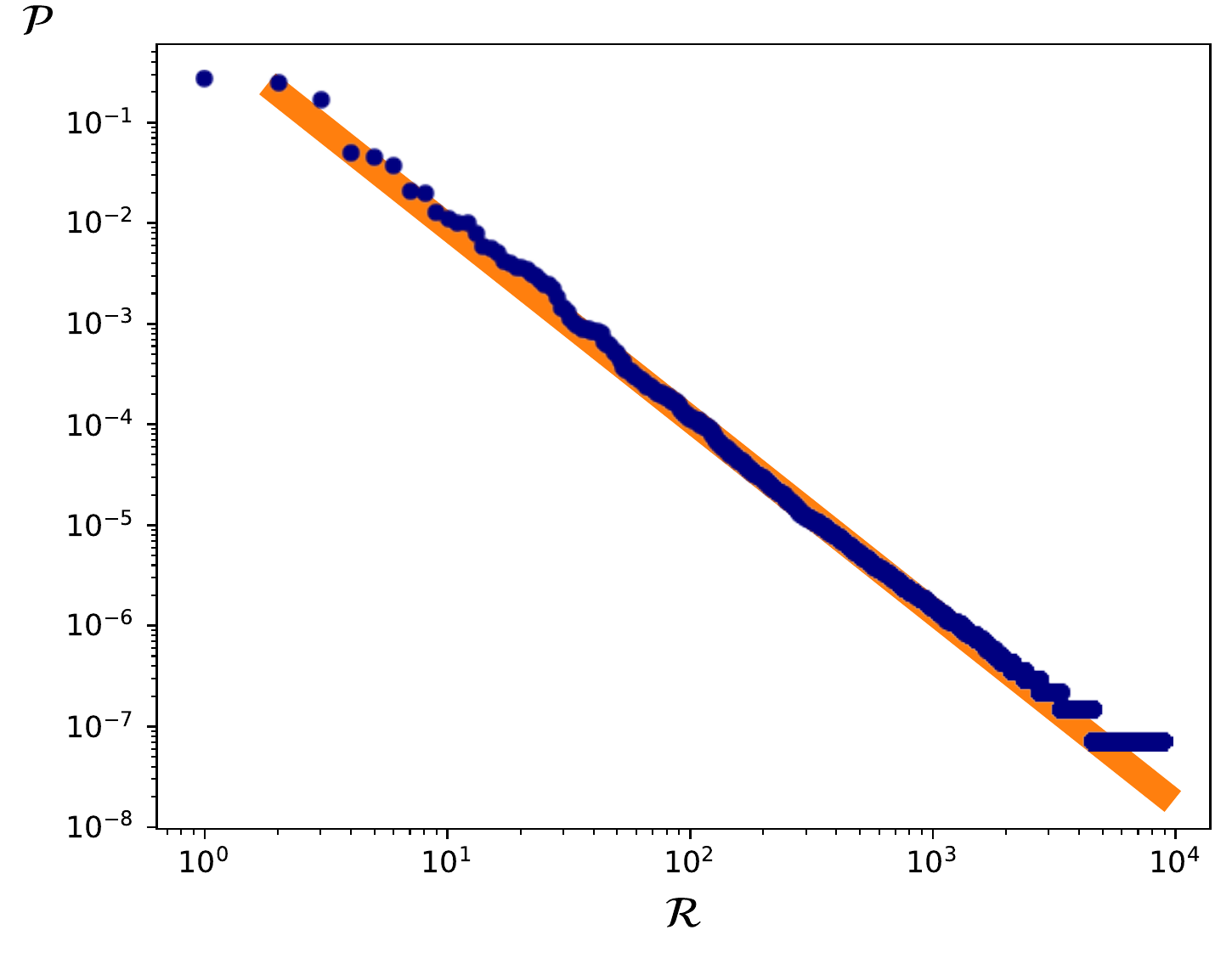}
\caption{\label{fig:zipf}Zipf's law \eqref{eq:Zipflaw} for the rank-frequency distribution of the continent sequences, with $\alpha\approx 1.9108$ with its uncertainty $u(\alpha)=0.0015$}
\end{figure}

The visualization of the latter as the rank frequency distribution is presented in \Cref{fig:zipf}.
The obtained diagram reveals the power-law dependence
\begin{equation}
\mathcal P(\mathcal R)\propto\mathcal R^{-\alpha},
\label{eq:Zipflaw}
\end{equation}
where $\mathcal P$ is the frequency of the sequence, $\mathcal R$ is the rank of this sequence and the exponent $\alpha\approx 1.9108$ with its uncertainty $u(\alpha)=0.0015$.
Such an inverse relation in the rank-frequency distribution is known as Zipf's law \cite{Zipf_1935,zipf_mejn}.

\begin{figure}[t]
\includegraphics[width=0.99\columnwidth]{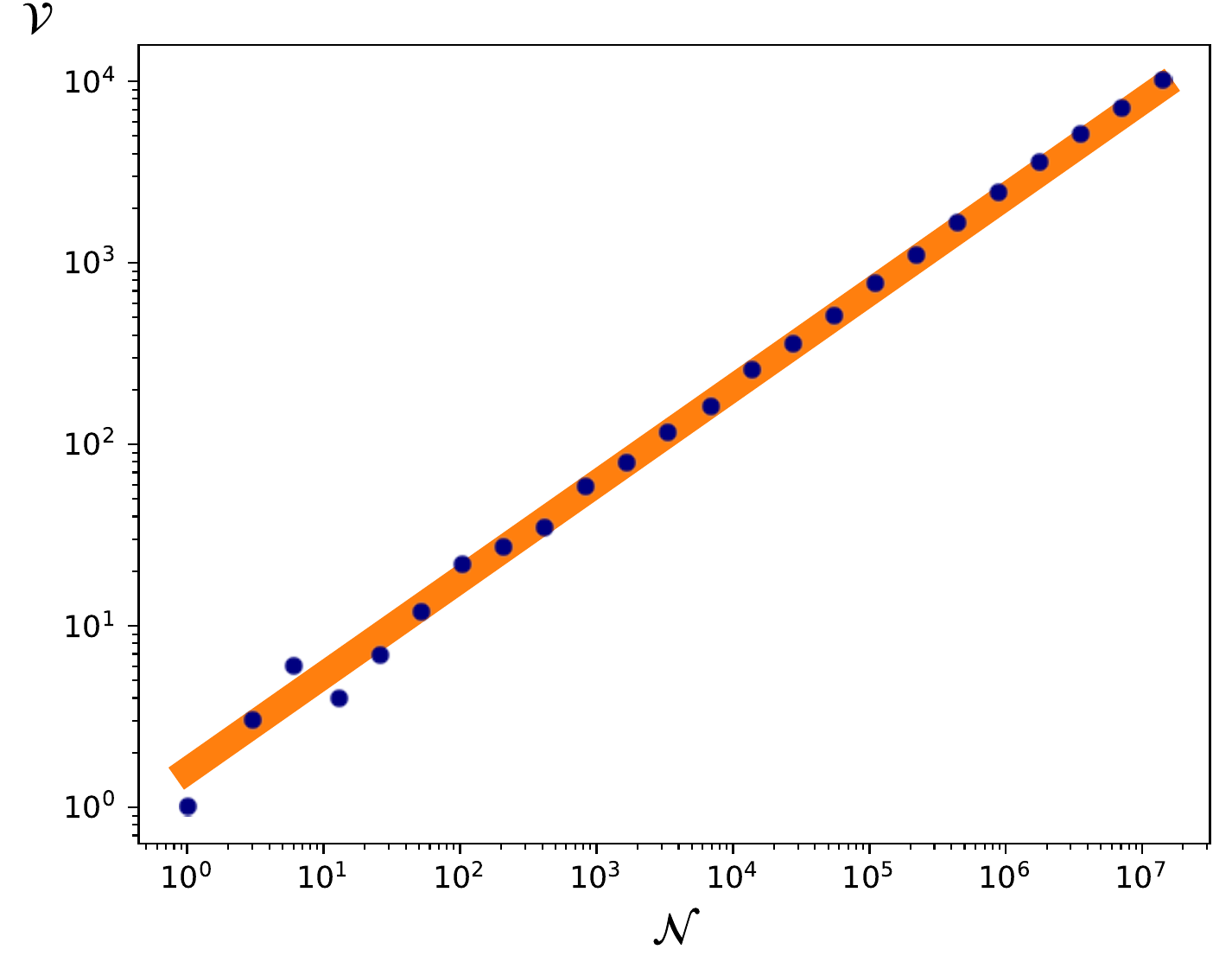}
\caption{\label{fig:heap}Heap's law \eqref{eq:Heapslaw} for the ``continent (number of countries)'' sequences, with $\beta\approx 0.527$ with its uncertainty $u(\beta)=0.014$}
\end{figure}

Since Zipf's law is often considered with Heap's law \cite{heap_lu,heap_fc}, defined as
\begin{equation}
    \mathcal V(\mathcal N)\propto \mathcal N^\beta,
    \label{eq:Heapslaw}
\end{equation}
where $\mathcal V$---in our case---is the number of distinct sequences ``continent (number of countries)'' in an $\mathcal N$ publications, we have also checked whether this law applies to our data. As can be read from \Cref{fig:heap}, where each point corresponds to a random sample of $\mathcal N$ articles, $\beta\approx 0.527(14)$ and as expected $\alpha=1/\beta$ \cite{heap_lu}.

\section{Discussion}

The most frequently observed sequences obtained from mapping authors' countries to continents do not show their intercontinental nature. 
The three most popular sequences (in the papers analyzed) correspond to single-country papers by authors affiliated in countries from North America ($\mathcal P\approx 26.51\%$), Europe ($\mathcal P\approx 24.59\%$) and Asia ($\mathcal P\approx 17.06\%$). 
The first case slightly deviates from the fitting to Zipf's law \eqref{eq:Zipflaw} as presented in \Cref{fig:zipf}.
The fourth place is taken by articles written by authors affiliated with institutes in two European countries ($\mathcal P\approx 4.95\%$).

The first intercontinental cooperation is ranked as the fifth and corresponds to papers written by authors from a single European country plus authors from a single country in North America. The abundance of such a sequence is $\mathcal P\approx 4.56\%$.
The next most frequent cases [Asia (1), North America (1) with $\mathcal P\approx 3.66\%$ and Asia (1), Europe (1) with $\mathcal P\approx 1.73\%$] correspond to a bilateral intercontinental cooperation established by authors from a pair of countries.
The seventh place is again taken by monocontinental cooperation between authors affiliated with countries from Australia \& Oceania.
In the eighth place of the ranking cooperation between one country from Asia and one country from Europe is placed.
The ninth place corresponds to the situation where authors are affiliated in two Asian countries.
Finally, the top-10 list is closed by papers committed by authors affiliated in two European countries and authors from one country (either USA or Canada) from North America.
The next three ranks do not reveal the intercontinental character of scientific cooperation (subsequently: with three European countries; two countries from North America; and single country from South America).
In contrast, the sequences on positions 14--20 reveal intercontinental cooperation.
 
The complete ranking containing nearly nine thousand sequences is available as a comma-separated-value file in Ref.~\onlinecite{data-intercont-rank}.

\section{Conclusions}

In this paper, the intercontinental cooperation among the authors of nearly 14 million articles collected from Scopus \cite{api} is studied. The obtained rank-frequency-based histogram reveals the power-law dependence \eqref{eq:Zipflaw}, which is known as Zipf's law, and occurs in many types of studied data---ranging from quantitative linguistics (where it was originally formulated \cite{Zipf_1935}), via social sciences \cite{cameron2022zipf}, music \cite{PEROTTI2020124309}, seismic activity \cite{zipf_vesp} to physics \cite{PhysRevE.98.032408}, popularity of chess openings \cite{PhysRevLett.103.218701}, and others \cite{zipf_mejn}.
The number of distinct alpha-numeric sequences ``continent (number of countries)'' scales with the number of analyzed papers according to the power law \eqref{eq:Heapslaw} with exponent $\beta\approx 1/\alpha$, which also recovers Heap's law---strongly correlated with Zipf's law.

The further studies may include detailed look into frequencies of sequences for fixed number continents (for example, for single, double, triples, etc., involved continents) or for more sophisticated methods of sequence analysis (by applying, for example, a scalar variable---as the inverse participation ratio---to distinguish among sequences). Also, a more thorough analysis of cooperation (based on international cooperation level) may exhibit some free-scaling behavior in rank-frequency distribution of analogous sequences but for countries. Detailed results of the studies mentioned above will be provided in the near future in a study under preparation by Krawczyk, Libirt and Malarz.

\begin{acknowledgments}
We are grateful to anonymous Referee for pointing out the possibility of analyzing our data not only on Zipf's but also on Heap's law.
\end{acknowledgments}

\bibliography{bib/scientometrics,bib/networks,bib/km,bib/this}

\begin{thebibliography}{55}%
\makeatletter
\providecommand \@ifxundefined [1]{%
 \@ifx{#1\undefined}
}%
\providecommand \@ifnum [1]{%
 \ifnum #1\expandafter \@firstoftwo
 \else \expandafter \@secondoftwo
 \fi
}%
\providecommand \@ifx [1]{%
 \ifx #1\expandafter \@firstoftwo
 \else \expandafter \@secondoftwo
 \fi
}%
\providecommand \natexlab [1]{#1}%
\providecommand \enquote  [1]{``#1''}%
\providecommand \bibnamefont  [1]{#1}%
\providecommand \bibfnamefont [1]{#1}%
\providecommand \citenamefont [1]{#1}%
\providecommand \href@noop [0]{\@secondoftwo}%
\providecommand \href [0]{\begingroup \@sanitize@url \@href}%
\providecommand \@href[1]{\@@startlink{#1}\@@href}%
\providecommand \@@href[1]{\endgroup#1\@@endlink}%
\providecommand \@sanitize@url [0]{\catcode `\\12\catcode `\$12\catcode
  `\&12\catcode `\#12\catcode `\^12\catcode `\_12\catcode `\%12\relax}%
\providecommand \@@startlink[1]{}%
\providecommand \@@endlink[0]{}%
\providecommand \url  [0]{\begingroup\@sanitize@url \@url }%
\providecommand \@url [1]{\endgroup\@href {#1}{\urlprefix }}%
\providecommand \urlprefix  [0]{URL }%
\providecommand \Eprint [0]{\href }%
\providecommand \doibase [0]{https://doi.org/}%
\providecommand \selectlanguage [0]{\@gobble}%
\providecommand \bibinfo  [0]{\@secondoftwo}%
\providecommand \bibfield  [0]{\@secondoftwo}%
\providecommand \translation [1]{[#1]}%
\providecommand \BibitemOpen [0]{}%
\providecommand \bibitemStop [0]{}%
\providecommand \bibitemNoStop [0]{.\EOS\space}%
\providecommand \EOS [0]{\spacefactor3000\relax}%
\providecommand \BibitemShut  [1]{\csname bibitem#1\endcsname}%
\let\auto@bib@innerbib\@empty
\bibitem [{\citenamefont {Melin}(2000)}]{WOS:000084641800003}%
  \BibitemOpen
  \bibfield  {author} {\bibinfo {author} {\bibfnamefont {G.}~\bibnamefont
  {Melin}},\ }\bibfield  {title} {\enquote {\bibinfo {title} {Pragmatism and
  self-organization---{R}esearch collaboration on the individual level},}\
  }\href {https://doi.org/10.1016/S0048-7333(99)00031-1} {\bibfield  {journal}
  {\bibinfo  {journal} {Research Policy}\ }\textbf {\bibinfo {volume} {29}},\
  \bibinfo {pages} {31--40} (\bibinfo {year} {2000})}\BibitemShut {NoStop}%
\bibitem [{\citenamefont {Wagner}\ and\ \citenamefont
  {Leydesdorff}(2005)}]{WOS:000233960500009}%
  \BibitemOpen
  \bibfield  {author} {\bibinfo {author} {\bibfnamefont {C.}~\bibnamefont
  {Wagner}}\ and\ \bibinfo {author} {\bibfnamefont {L.}~\bibnamefont
  {Leydesdorff}},\ }\bibfield  {title} {\enquote {\bibinfo {title} {Network
  structure, self-organization, and the growth of international collaboration
  in science},}\ }\href {https://doi.org/10.1016/j.respol.2005.08.002}
  {\bibfield  {journal} {\bibinfo  {journal} {Research Policy}\ }\textbf
  {\bibinfo {volume} {34}},\ \bibinfo {pages} {1608--1618} (\bibinfo {year}
  {2005})}\BibitemShut {NoStop}%
\bibitem [{\citenamefont {Leydesdorff}\ and\ \citenamefont
  {Wagner}(2008)}]{WOS:000260903700007}%
  \BibitemOpen
  \bibfield  {author} {\bibinfo {author} {\bibfnamefont {L.}~\bibnamefont
  {Leydesdorff}}\ and\ \bibinfo {author} {\bibfnamefont {C.~S.}\ \bibnamefont
  {Wagner}},\ }\bibfield  {title} {\enquote {\bibinfo {title} {International
  collaboration in science and the formation of a core group},}\ }\href
  {https://doi.org/10.1016/j.joi.2008.07.003} {\bibfield  {journal} {\bibinfo
  {journal} {Journal of Informetrics}\ }\textbf {\bibinfo {volume} {2}},\
  \bibinfo {pages} {317--325} (\bibinfo {year} {2008})}\BibitemShut {NoStop}%
\bibitem [{\citenamefont {Luukkonen}, \citenamefont {Persson},\ and\
  \citenamefont {Sivertsen}(1992)}]{WOS:A1992GW95200006}%
  \BibitemOpen
  \bibfield  {author} {\bibinfo {author} {\bibfnamefont {T.}~\bibnamefont
  {Luukkonen}}, \bibinfo {author} {\bibfnamefont {O.}~\bibnamefont {Persson}},\
  and\ \bibinfo {author} {\bibfnamefont {G.}~\bibnamefont {Sivertsen}},\
  }\bibfield  {title} {\enquote {\bibinfo {title} {Understanding patterns of
  international scientific collaboration},}\ }\href
  {https://doi.org/10.1177/016224399201700106} {\bibfield  {journal} {\bibinfo
  {journal} {Science Technology \& Human Values}\ }\textbf {\bibinfo {volume}
  {17}},\ \bibinfo {pages} {101--126} (\bibinfo {year} {1992})}\BibitemShut
  {NoStop}%
\bibitem [{\citenamefont {Melin}\ and\ \citenamefont
  {Persson}(1996)}]{WOS:A1996VG50500006}%
  \BibitemOpen
  \bibfield  {author} {\bibinfo {author} {\bibfnamefont {G.}~\bibnamefont
  {Melin}}\ and\ \bibinfo {author} {\bibfnamefont {O.}~\bibnamefont
  {Persson}},\ }\bibfield  {title} {\enquote {\bibinfo {title} {Studying
  research collaboration using co-authorships},}\ }\href
  {https://doi.org/10.1007/BF02129600} {\bibfield  {journal} {\bibinfo
  {journal} {Scientometrics}\ }\textbf {\bibinfo {volume} {36}},\ \bibinfo
  {pages} {363--377} (\bibinfo {year} {1996})}\BibitemShut {NoStop}%
\bibitem [{\citenamefont {Katz}\ and\ \citenamefont
  {Martin}(1997)}]{WOS:A1997XC57700001}%
  \BibitemOpen
  \bibfield  {author} {\bibinfo {author} {\bibfnamefont {J.}~\bibnamefont
  {Katz}}\ and\ \bibinfo {author} {\bibfnamefont {B.}~\bibnamefont {Martin}},\
  }\bibfield  {title} {\enquote {\bibinfo {title} {What is research
  collaboration?}}\ }\href {https://doi.org/10.1016/S0048-7333(96)00917-1}
  {\bibfield  {journal} {\bibinfo  {journal} {Research Policy}\ }\textbf
  {\bibinfo {volume} {26}},\ \bibinfo {pages} {1--18} (\bibinfo {year}
  {1997})}\BibitemShut {NoStop}%
\bibitem [{\citenamefont {Gazni}, \citenamefont {Sugimoto},\ and\ \citenamefont
  {Didegah}(2012)}]{WOS:000302157900009}%
  \BibitemOpen
  \bibfield  {author} {\bibinfo {author} {\bibfnamefont {A.}~\bibnamefont
  {Gazni}}, \bibinfo {author} {\bibfnamefont {C.~R.}\ \bibnamefont
  {Sugimoto}},\ and\ \bibinfo {author} {\bibfnamefont {F.}~\bibnamefont
  {Didegah}},\ }\bibfield  {title} {\enquote {\bibinfo {title} {Mapping world
  scientific collaboration: {A}uthors, institutions, and countries},}\ }\href
  {https://doi.org/10.1002/asi.21688} {\bibfield  {journal} {\bibinfo
  {journal} {Journal of the American Society for Information Science and
  Technology}\ }\textbf {\bibinfo {volume} {63}},\ \bibinfo {pages} {323--335}
  (\bibinfo {year} {2012})}\BibitemShut {NoStop}%
\bibitem [{\citenamefont {Gl\"anzel}(2001)}]{WOS:000169909700005}%
  \BibitemOpen
  \bibfield  {author} {\bibinfo {author} {\bibfnamefont {W.}~\bibnamefont
  {Gl\"anzel}},\ }\bibfield  {title} {\enquote {\bibinfo {title} {National
  characteristics in international scientific co-authorship relations},}\
  }\href {https://doi.org/10.1023/A:1010512628145} {\bibfield  {journal}
  {\bibinfo  {journal} {Scientometrics}\ }\textbf {\bibinfo {volume} {51}},\
  \bibinfo {pages} {69--115} (\bibinfo {year} {2001})}\BibitemShut {NoStop}%
\bibitem [{\citenamefont {Pond}, \citenamefont {van Oort},\ and\ \citenamefont
  {Frenken}(2007)}]{WOS:000249187700005}%
  \BibitemOpen
  \bibfield  {author} {\bibinfo {author} {\bibfnamefont {R.}~\bibnamefont
  {Pond}}, \bibinfo {author} {\bibfnamefont {F.}~\bibnamefont {van Oort}},\
  and\ \bibinfo {author} {\bibfnamefont {K.}~\bibnamefont {Frenken}},\
  }\bibfield  {title} {\enquote {\bibinfo {title} {The geographical and
  institutional proximity of research collaboration},}\ }\href
  {https://doi.org/10.1111/j.1435-5957.2007.00126.x} {\bibfield  {journal}
  {\bibinfo  {journal} {Papers in Regional Science}\ }\textbf {\bibinfo
  {volume} {86}},\ \bibinfo {pages} {423--443} (\bibinfo {year}
  {2007})}\BibitemShut {NoStop}%
\bibitem [{\citenamefont {Luukkonen}\ \emph {et~al.}(1993)\citenamefont
  {Luukkonen}, \citenamefont {Tijssen}, \citenamefont {Persson},\ and\
  \citenamefont {Sivertsen}}]{WOS:A1993LW52100002}%
  \BibitemOpen
  \bibfield  {author} {\bibinfo {author} {\bibfnamefont {T.}~\bibnamefont
  {Luukkonen}}, \bibinfo {author} {\bibfnamefont {R.}~\bibnamefont {Tijssen}},
  \bibinfo {author} {\bibfnamefont {O.}~\bibnamefont {Persson}},\ and\ \bibinfo
  {author} {\bibfnamefont {G.}~\bibnamefont {Sivertsen}},\ }\bibfield  {title}
  {\enquote {\bibinfo {title} {The measurement of international scientific
  collaboration},}\ }\href {https://doi.org/10.1007/BF02016282} {\bibfield
  {journal} {\bibinfo  {journal} {Scientometrics}\ }\textbf {\bibinfo {volume}
  {28}},\ \bibinfo {pages} {15--36} (\bibinfo {year} {1993})}\BibitemShut
  {NoStop}%
\bibitem [{WOS(2023)}]{WOS}%
  \BibitemOpen
  \href {https://www.webofknowledge.com/} {\enquote {\bibinfo {title} {Web of
  {S}cience, https://www.webofknowledge.com/},}\ } (\bibinfo {year}
  {2023})\BibitemShut {NoStop}%
\bibitem [{\citenamefont {Kozma}\ and\ \citenamefont
  {Calero-Medina}(2019)}]{Kozma_2019}%
  \BibitemOpen
  \bibfield  {author} {\bibinfo {author} {\bibfnamefont {C.}~\bibnamefont
  {Kozma}}\ and\ \bibinfo {author} {\bibfnamefont {C.}~\bibnamefont
  {Calero-Medina}},\ }\bibfield  {title} {\enquote {\bibinfo {title} {The role
  of {S}outh {A}frican researchers in intercontinental collaboration},}\ }\href
  {https://doi.org/10.1007/s11192-019-03230-9} {\bibfield  {journal} {\bibinfo
  {journal} {Scientometrics}\ }\textbf {\bibinfo {volume} {121}},\ \bibinfo
  {pages} {1293--1321} (\bibinfo {year} {2019})}\BibitemShut {NoStop}%
\bibitem [{\citenamefont {Abramo}, \citenamefont {D'Angelo},\ and\
  \citenamefont {Di~Costa}(2020)}]{Abramo_2020}%
  \BibitemOpen
  \bibfield  {author} {\bibinfo {author} {\bibfnamefont {G.}~\bibnamefont
  {Abramo}}, \bibinfo {author} {\bibfnamefont {C.~A.}\ \bibnamefont
  {D'Angelo}},\ and\ \bibinfo {author} {\bibfnamefont {F.}~\bibnamefont
  {Di~Costa}},\ }\bibfield  {title} {\enquote {\bibinfo {title} {The role of
  geographical proximity in knowledge diffusion, measured by citations to
  scientific literature},}\ }\href {https://doi.org/10.1016/j.joi.2020.101010}
  {\bibfield  {journal} {\bibinfo  {journal} {Journal of Informetrics}\
  }\textbf {\bibinfo {volume} {14}},\ \bibinfo {pages} {101010} (\bibinfo
  {year} {2020})}\BibitemShut {NoStop}%
\bibitem [{\citenamefont {Zipf}(1935)}]{Zipf_1935}%
  \BibitemOpen
  \bibfield  {author} {\bibinfo {author} {\bibfnamefont {G.~K.}\ \bibnamefont
  {Zipf}},\ }\href@noop {} {\emph {\bibinfo {title} {The Psycho-biology of
  Language: {A}n Introduction to Dynamic Philology}}}\ (\bibinfo  {publisher}
  {Houghton Mifflin},\ \bibinfo {year} {1935})\BibitemShut {NoStop}%
\bibitem [{api(2022)}]{api}%
  \BibitemOpen
  \href {http://api.elsevier.com/content/search/scopus} {\enquote {\bibinfo
  {title} {Scopus, http://api.elsevier.com/content/search/scopus},}\ }
  (\bibinfo {year} {2022})\BibitemShut {NoStop}%
\bibitem [{WoS(2023{\natexlab{a}})}]{WoS_Vespignani}%
  \BibitemOpen
  \href {https://www.webofscience.com/wos/author/record/614437} {\enquote
  {\bibinfo {title} {Alessandro {V}espignani, {W}eb of {S}cience
  {R}esearcher{ID}: A-8794-2009},}\ } (\bibinfo {year} {September 4,
  2023}{\natexlab{a}})\BibitemShut {NoStop}%
\bibitem [{\citenamefont {Hofman}\ \emph {et~al.}(2021)\citenamefont {Hofman}
  \emph {et~al.}}]{WOS:000668492600013}%
  \BibitemOpen
  \bibfield  {author} {\bibinfo {author} {\bibfnamefont {J.~M.}\ \bibnamefont
  {Hofman}} \emph {et~al.},\ }\bibfield  {title} {\enquote {\bibinfo {title}
  {Integrating explanation and prediction in computational social science},}\
  }\href {https://doi.org/10.1038/s41586-021-03659-0} {\bibfield  {journal}
  {\bibinfo  {journal} {Nature}\ }\textbf {\bibinfo {volume} {595}},\ \bibinfo
  {pages} {181--188} (\bibinfo {year} {2021})}\BibitemShut {NoStop}%
\bibitem [{\citenamefont {Lazer}\ \emph {et~al.}(2020)\citenamefont {Lazer}
  \emph {et~al.}}]{WOS:000567522200024}%
  \BibitemOpen
  \bibfield  {author} {\bibinfo {author} {\bibfnamefont {D.~M.~J.}\
  \bibnamefont {Lazer}} \emph {et~al.},\ }\bibfield  {title} {\enquote
  {\bibinfo {title} {Computational social science: {O}bstacles and
  opportunities},}\ }\href {https://doi.org/10.1126/science.aaz8170} {\bibfield
   {journal} {\bibinfo  {journal} {Science}\ }\textbf {\bibinfo {volume}
  {369}},\ \bibinfo {pages} {1060--1062} (\bibinfo {year} {2020})}\BibitemShut
  {NoStop}%
\bibitem [{\citenamefont {Zhang}, \citenamefont {Karsai},\ and\ \citenamefont
  {Vespignani}(2018)}]{WOS:000444788000001}%
  \BibitemOpen
  \bibfield  {author} {\bibinfo {author} {\bibfnamefont {Q.}~\bibnamefont
  {Zhang}}, \bibinfo {author} {\bibfnamefont {M.}~\bibnamefont {Karsai}},\ and\
  \bibinfo {author} {\bibfnamefont {A.}~\bibnamefont {Vespignani}},\ }\bibfield
   {title} {\enquote {\bibinfo {title} {Link transmission centrality in
  large-scale social networks},}\ }\href
  {https://doi.org/10.1140/epjds/s13688-018-0162-8} {\bibfield  {journal}
  {\bibinfo  {journal} {EPJ Data Science}\ }\textbf {\bibinfo {volume} {7}},\
  \bibinfo {pages} {33} (\bibinfo {year} {2018})}\BibitemShut {NoStop}%
\bibitem [{\citenamefont {Colizza}\ \emph {et~al.}(2005)\citenamefont
  {Colizza}, \citenamefont {Flammini}, \citenamefont {Maritan},\ and\
  \citenamefont {Vespignani}}]{WOS:000229193300002}%
  \BibitemOpen
  \bibfield  {author} {\bibinfo {author} {\bibfnamefont {V.}~\bibnamefont
  {Colizza}}, \bibinfo {author} {\bibfnamefont {A.}~\bibnamefont {Flammini}},
  \bibinfo {author} {\bibfnamefont {A.}~\bibnamefont {Maritan}},\ and\ \bibinfo
  {author} {\bibfnamefont {A.}~\bibnamefont {Vespignani}},\ }\bibfield  {title}
  {\enquote {\bibinfo {title} {Characterization and modeling of protein-protein
  interaction networks},}\ }\href {https://doi.org/10.1016/j.physa.2004.12.030}
  {\bibfield  {journal} {\bibinfo  {journal} {Physica A---statistical Mechanics
  and its Applications}\ }\textbf {\bibinfo {volume} {352}},\ \bibinfo {pages}
  {1--27} (\bibinfo {year} {2005})}\BibitemShut {NoStop}%
\bibitem [{\citenamefont {Moreno}, \citenamefont {Pastor-Satorras},\ and\
  \citenamefont {Vespignani}(2002)}]{WOS:000175859600017}%
  \BibitemOpen
  \bibfield  {author} {\bibinfo {author} {\bibfnamefont {Y.}~\bibnamefont
  {Moreno}}, \bibinfo {author} {\bibfnamefont {R.}~\bibnamefont
  {Pastor-Satorras}},\ and\ \bibinfo {author} {\bibfnamefont {A.}~\bibnamefont
  {Vespignani}},\ }\bibfield  {title} {\enquote {\bibinfo {title} {Epidemic
  outbreaks in complex heterogeneous networks},}\ }\href
  {https://doi.org/10.1140/epjb/e20020122} {\bibfield  {journal} {\bibinfo
  {journal} {European Physical Journal B}\ }\textbf {\bibinfo {volume} {26}},\
  \bibinfo {pages} {521--529} (\bibinfo {year} {2002})}\BibitemShut {NoStop}%
\bibitem [{\citenamefont {Barrat}\ \emph {et~al.}(2004)\citenamefont {Barrat},
  \citenamefont {Barthelemy}, \citenamefont {Pastor-Satorras},\ and\
  \citenamefont {Vespignani}}]{WOS:000220314500008}%
  \BibitemOpen
  \bibfield  {author} {\bibinfo {author} {\bibfnamefont {A.}~\bibnamefont
  {Barrat}}, \bibinfo {author} {\bibfnamefont {M.}~\bibnamefont {Barthelemy}},
  \bibinfo {author} {\bibfnamefont {R.}~\bibnamefont {Pastor-Satorras}},\ and\
  \bibinfo {author} {\bibfnamefont {A.}~\bibnamefont {Vespignani}},\ }\bibfield
   {title} {\enquote {\bibinfo {title} {The architecture of complex weighted
  networks},}\ }\href {https://doi.org/10.1073/pnas.0400087101} {\bibfield
  {journal} {\bibinfo  {journal} {Proceedings of the National Academy of
  Sciences of the United States of America}\ }\textbf {\bibinfo {volume}
  {101}},\ \bibinfo {pages} {3747--3752} (\bibinfo {year} {2004})}\BibitemShut
  {NoStop}%
\bibitem [{\citenamefont {Colizza}\ \emph {et~al.}(2006)\citenamefont
  {Colizza}, \citenamefont {Flammini}, \citenamefont {Serrano},\ and\
  \citenamefont {Vespignani}}]{WOS:000235464700021}%
  \BibitemOpen
  \bibfield  {author} {\bibinfo {author} {\bibfnamefont {V.}~\bibnamefont
  {Colizza}}, \bibinfo {author} {\bibfnamefont {A.}~\bibnamefont {Flammini}},
  \bibinfo {author} {\bibfnamefont {M.}~\bibnamefont {Serrano}},\ and\ \bibinfo
  {author} {\bibfnamefont {A.}~\bibnamefont {Vespignani}},\ }\bibfield  {title}
  {\enquote {\bibinfo {title} {Detecting rich-club ordering in complex
  networks},}\ }\href {https://doi.org/10.1038/nphys209} {\bibfield  {journal}
  {\bibinfo  {journal} {Nature Physics}\ }\textbf {\bibinfo {volume} {2}},\
  \bibinfo {pages} {110--115} (\bibinfo {year} {2006})}\BibitemShut {NoStop}%
\bibitem [{\citenamefont {Cramer}\ \emph {et~al.}(2022)\citenamefont {Cramer}
  \emph {et~al.}}]{WOS:000819659900005_at_all}%
  \BibitemOpen
  \bibfield  {author} {\bibinfo {author} {\bibfnamefont {E.~Y.}\ \bibnamefont
  {Cramer}} \emph {et~al.},\ }\bibfield  {title} {\enquote {\bibinfo {title}
  {Evaluation of individual and ensemble probabilistic forecasts of {COVID}-19
  mortality in the {U}nited {S}tates},}\ }\href
  {https://doi.org/10.1073/pnas.2113561119} {\bibfield  {journal} {\bibinfo
  {journal} {Proceedings of the National Academy of Sciences of the United
  States of America}\ }\textbf {\bibinfo {volume} {119}},\ \bibinfo {pages}
  {e2113561119} (\bibinfo {year} {2022})}\BibitemShut {NoStop}%
\bibitem [{\citenamefont {Reich}\ \emph {et~al.}(2022)\citenamefont {Reich}
  \emph {et~al.}}]{WOS:000805653200014}%
  \BibitemOpen
  \bibfield  {author} {\bibinfo {author} {\bibfnamefont {N.~G.}\ \bibnamefont
  {Reich}} \emph {et~al.},\ }\bibfield  {title} {\enquote {\bibinfo {title}
  {Collaborative hubs: {M}aking the most of predictive epidemic modeling},}\
  }\href {https://doi.org/10.2105/AJPH.2022.306831} {\bibfield  {journal}
  {\bibinfo  {journal} {American Journal of Public Health}\ }\textbf {\bibinfo
  {volume} {112}},\ \bibinfo {pages} {839--842} (\bibinfo {year}
  {2022})}\BibitemShut {NoStop}%
\bibitem [{\citenamefont {Truelove}\ \emph {et~al.}(2022)\citenamefont
  {Truelove} \emph {et~al.}}]{WOS:000815971000001_et_all}%
  \BibitemOpen
  \bibfield  {author} {\bibinfo {author} {\bibfnamefont {S.}~\bibnamefont
  {Truelove}} \emph {et~al.},\ }\bibfield  {title} {\enquote {\bibinfo {title}
  {Projected resurgence of {COVID}-19 in the {U}nited {S}tates in
  {J}uly-{D}ecember 2021 resulting from the increased transmissibility of the
  {D}elta variant and faltering vaccination},}\ }\href
  {https://doi.org/10.7554/eLife.73584} {\bibfield  {journal} {\bibinfo
  {journal} {eLife}\ }\textbf {\bibinfo {volume} {11}},\ \bibinfo {pages}
  {e73584} (\bibinfo {year} {2022})}\BibitemShut {NoStop}%
\bibitem [{\citenamefont {Liu}\ \emph {et~al.}(2022)\citenamefont {Liu},
  \citenamefont {Zhang}, \citenamefont {Peng}, \citenamefont {Litvinova},
  \citenamefont {Huang}, \citenamefont {Poletti}, \citenamefont {Trentini},
  \citenamefont {Guzzetta}, \citenamefont {Marziano}, \citenamefont {Zhou},
  \citenamefont {Viboud}, \citenamefont {Bento}, \citenamefont {Lv},
  \citenamefont {Vespignani}, \citenamefont {Merler}, \citenamefont {Yu},\ and\
  \citenamefont {Ajelli}}]{WOS:000742728600007}%
  \BibitemOpen
  \bibfield  {author} {\bibinfo {author} {\bibfnamefont {Q.-H.}\ \bibnamefont
  {Liu}}, \bibinfo {author} {\bibfnamefont {J.}~\bibnamefont {Zhang}}, \bibinfo
  {author} {\bibfnamefont {C.}~\bibnamefont {Peng}}, \bibinfo {author}
  {\bibfnamefont {M.}~\bibnamefont {Litvinova}}, \bibinfo {author}
  {\bibfnamefont {S.}~\bibnamefont {Huang}}, \bibinfo {author} {\bibfnamefont
  {P.}~\bibnamefont {Poletti}}, \bibinfo {author} {\bibfnamefont
  {F.}~\bibnamefont {Trentini}}, \bibinfo {author} {\bibfnamefont
  {G.}~\bibnamefont {Guzzetta}}, \bibinfo {author} {\bibfnamefont
  {V.}~\bibnamefont {Marziano}}, \bibinfo {author} {\bibfnamefont
  {T.}~\bibnamefont {Zhou}}, \bibinfo {author} {\bibfnamefont {C.}~\bibnamefont
  {Viboud}}, \bibinfo {author} {\bibfnamefont {A.~I.}\ \bibnamefont {Bento}},
  \bibinfo {author} {\bibfnamefont {J.}~\bibnamefont {Lv}}, \bibinfo {author}
  {\bibfnamefont {A.}~\bibnamefont {Vespignani}}, \bibinfo {author}
  {\bibfnamefont {S.}~\bibnamefont {Merler}}, \bibinfo {author} {\bibfnamefont
  {H.}~\bibnamefont {Yu}},\ and\ \bibinfo {author} {\bibfnamefont
  {M.}~\bibnamefont {Ajelli}},\ }\bibfield  {title} {\enquote {\bibinfo {title}
  {Model-based evaluation of alternative reactive class closure strategies
  against {COVID}-19},}\ }\href {https://doi.org/10.1038/s41467-021-27939-5}
  {\bibfield  {journal} {\bibinfo  {journal} {Nature Communications}\ }\textbf
  {\bibinfo {volume} {13}},\ \bibinfo {pages} {322} (\bibinfo {year}
  {2022})}\BibitemShut {NoStop}%
\bibitem [{\citenamefont {Davis}\ \emph {et~al.}(2021)\citenamefont {Davis},
  \citenamefont {Chinazzi}, \citenamefont {Perra}, \citenamefont {Mu},
  \citenamefont {Piontti}, \citenamefont {Ajelli}, \citenamefont {Dean},
  \citenamefont {Gioannini}, \citenamefont {Litvinova}, \citenamefont {Merler},
  \citenamefont {Rossi}, \citenamefont {Sun}, \citenamefont {Xiong},
  \citenamefont {Longini}, \citenamefont {Halloran}, \citenamefont {Viboud},\
  and\ \citenamefont {Vespignani}}]{WOS:000722150700001}%
  \BibitemOpen
  \bibfield  {author} {\bibinfo {author} {\bibfnamefont {J.~T.}\ \bibnamefont
  {Davis}}, \bibinfo {author} {\bibfnamefont {M.}~\bibnamefont {Chinazzi}},
  \bibinfo {author} {\bibfnamefont {N.}~\bibnamefont {Perra}}, \bibinfo
  {author} {\bibfnamefont {K.}~\bibnamefont {Mu}}, \bibinfo {author}
  {\bibfnamefont {A.}~\bibnamefont {Piontti}}, \bibinfo {author} {\bibfnamefont
  {M.}~\bibnamefont {Ajelli}}, \bibinfo {author} {\bibfnamefont {N.~E.}\
  \bibnamefont {Dean}}, \bibinfo {author} {\bibfnamefont {C.}~\bibnamefont
  {Gioannini}}, \bibinfo {author} {\bibfnamefont {M.}~\bibnamefont
  {Litvinova}}, \bibinfo {author} {\bibfnamefont {S.}~\bibnamefont {Merler}},
  \bibinfo {author} {\bibfnamefont {L.}~\bibnamefont {Rossi}}, \bibinfo
  {author} {\bibfnamefont {K.}~\bibnamefont {Sun}}, \bibinfo {author}
  {\bibfnamefont {X.}~\bibnamefont {Xiong}}, \bibinfo {author} {\bibfnamefont
  {I.~M.}\ \bibnamefont {Longini}, \bibfnamefont {Jr.}}, \bibinfo {author}
  {\bibfnamefont {M.~E.}\ \bibnamefont {Halloran}}, \bibinfo {author}
  {\bibfnamefont {C.}~\bibnamefont {Viboud}},\ and\ \bibinfo {author}
  {\bibfnamefont {A.}~\bibnamefont {Vespignani}},\ }\bibfield  {title}
  {\enquote {\bibinfo {title} {Cryptic transmission of {SARS}-{CoV}-2 and the
  first {COVID}-19 wave},}\ }\href {https://doi.org/10.1038/s41586-021-04130-w}
  {\bibfield  {journal} {\bibinfo  {journal} {Nature}\ }\textbf {\bibinfo
  {volume} {600}},\ \bibinfo {pages} {127–132} (\bibinfo {year}
  {2021})}\BibitemShut {NoStop}%
\bibitem [{ccs(2023)}]{ccs-soc}%
  \BibitemOpen
  \href {https://cssociety.org/community/awards} {\enquote {\bibinfo {title}
  {https://cssociety.org/community/awards},}\ } (\bibinfo {year}
  {2023})\BibitemShut {NoStop}%
\bibitem [{WoS(2023{\natexlab{b}})}]{WoS_Stanley}%
  \BibitemOpen
  \href {https://www.webofscience.com/wos/author/record/29989377} {\enquote
  {\bibinfo {title} {H. {E}ugene {S}tanley, {W}eb of {S}cience
  {R}esearcher{ID}: {GEF}-8992-2022},}\ } (\bibinfo {year} {September 4,
  2023}{\natexlab{b}})\BibitemShut {NoStop}%
\bibitem [{WoS(2023{\natexlab{c}})}]{WoS_Miguel}%
  \BibitemOpen
  \href {https://www.webofscience.com/wos/author/record/1824277} {\enquote
  {\bibinfo {title} {Maxi {S}an {M}iguel, {W}eb of {S}cience {R}esearcher{ID}:
  G-3224-2015},}\ } (\bibinfo {year} {September 4,
  2023}{\natexlab{c}})\BibitemShut {NoStop}%
\bibitem [{WoS(2023{\natexlab{d}})}]{WoS_Batty}%
  \BibitemOpen
  \href {https://www.webofscience.com/wos/author/record/292055} {\enquote
  {\bibinfo {title} {Michael {B}atty, {W}eb of {S}cience {R}esearcher{ID}:
  D-4422-2011},}\ } (\bibinfo {year} {September 4,
  2023}{\natexlab{d}})\BibitemShut {NoStop}%
\bibitem [{WoS(2023{\natexlab{e}})}]{WoS_Barabasi}%
  \BibitemOpen
  \href {https://www.webofscience.com/wos/author/record/1370580} {\enquote
  {\bibinfo {title} {Albert-{L}ászló {B}arabási, {W}eb of {S}cience
  {R}esearcher{ID}: S-6474-2017},}\ } (\bibinfo {year} {September 4,
  2023}{\natexlab{e}})\BibitemShut {NoStop}%
\bibitem [{WoS(2023{\natexlab{f}})}]{WoS_Moreno}%
  \BibitemOpen
  \href {https://www.webofscience.com/wos/author/record/941862} {\enquote
  {\bibinfo {title} {Yamir {M}oreno, {W}eb of {S}cience {R}esearcher{ID}:
  A-1076-2009},}\ } (\bibinfo {year} {September 4,
  2023}{\natexlab{f}})\BibitemShut {NoStop}%
\bibitem [{WoS(2023{\natexlab{g}})}]{WoS_Mendes}%
  \BibitemOpen
  \href {https://www.webofscience.com/wos/author/record/333767} {\enquote
  {\bibinfo {title} {Jose {F}ernando {F}erreira {M}endes, {W}eb of {S}cience
  {R}esearcher{ID}: A-6996-2010},}\ } (\bibinfo {year} {September 4,
  2023}{\natexlab{g}})\BibitemShut {NoStop}%
\bibitem [{\citenamefont {Vespignani}\ \emph {et~al.}(2020)\citenamefont
  {Vespignani}, \citenamefont {Tian}, \citenamefont {Dye}, \citenamefont
  {Lloyd-Smith}, \citenamefont {Eggo}, \citenamefont {Shrestha}, \citenamefont
  {Scarpino}, \citenamefont {Gutierrez}, \citenamefont {Kraemer}, \citenamefont
  {Wu}, \citenamefont {Leung},\ and\ \citenamefont {Leung}}]{cc}%
  \BibitemOpen
  \bibfield  {author} {\bibinfo {author} {\bibfnamefont {A.}~\bibnamefont
  {Vespignani}}, \bibinfo {author} {\bibfnamefont {H.}~\bibnamefont {Tian}},
  \bibinfo {author} {\bibfnamefont {C.}~\bibnamefont {Dye}}, \bibinfo {author}
  {\bibfnamefont {J.}~\bibnamefont {Lloyd-Smith}}, \bibinfo {author}
  {\bibfnamefont {R.}~\bibnamefont {Eggo}}, \bibinfo {author} {\bibfnamefont
  {M.}~\bibnamefont {Shrestha}}, \bibinfo {author} {\bibfnamefont
  {S.}~\bibnamefont {Scarpino}}, \bibinfo {author} {\bibfnamefont
  {B.}~\bibnamefont {Gutierrez}}, \bibinfo {author} {\bibfnamefont
  {M.}~\bibnamefont {Kraemer}}, \bibinfo {author} {\bibfnamefont
  {J.}~\bibnamefont {Wu}}, \bibinfo {author} {\bibfnamefont {K.}~\bibnamefont
  {Leung}},\ and\ \bibinfo {author} {\bibfnamefont {G.}~\bibnamefont {Leung}},\
  }\bibfield  {title} {\enquote {\bibinfo {title} {Modelling {COVID}-19},}\
  }\href {https://doi.org/10.1038/s42254-020-0178-4} {\bibfield  {journal}
  {\bibinfo  {journal} {Nature Reviews Physics}\ }\textbf {\bibinfo {volume}
  {2}},\ \bibinfo {pages} {279--281} (\bibinfo {year} {2020})}\BibitemShut
  {NoStop}%
\bibitem [{\citenamefont {Aarts}\ \emph {et~al.}(2017)\citenamefont {Aarts}
  \emph {et~al.}}]{example_6_at_all}%
  \BibitemOpen
  \bibfield  {author} {\bibinfo {author} {\bibfnamefont {G.}~\bibnamefont
  {Aarts}} \emph {et~al.},\ }\bibfield  {title} {\enquote {\bibinfo {title}
  {Heavy-flavor production and medium properties in high-energy nuclear
  collisions---{W}hat next?}}\ }\href
  {https://doi.org/10.1140/epja/i2017-12282-9} {\bibfield  {journal} {\bibinfo
  {journal} {The European Physical Journal A}\ }\textbf {\bibinfo {volume}
  {53}},\ \bibinfo {pages} {93} (\bibinfo {year} {2017})}\BibitemShut {NoStop}%
\bibitem [{\citenamefont {Krawczyk}\ \emph {et~al.}(2005)\citenamefont
  {Krawczyk}, \citenamefont {Malarz}, \citenamefont {Kawecka-Magiera},
  \citenamefont {Maksymowicz},\ and\ \citenamefont
  {Ku{\l}akowski}}]{Krawczyk2005}%
  \BibitemOpen
  \bibfield  {author} {\bibinfo {author} {\bibfnamefont {M.~J.}\ \bibnamefont
  {Krawczyk}}, \bibinfo {author} {\bibfnamefont {K.}~\bibnamefont {Malarz}},
  \bibinfo {author} {\bibfnamefont {B.}~\bibnamefont {Kawecka-Magiera}},
  \bibinfo {author} {\bibfnamefont {A.~Z.}\ \bibnamefont {Maksymowicz}},\ and\
  \bibinfo {author} {\bibfnamefont {K.}~\bibnamefont {Ku{\l}akowski}},\
  }\bibfield  {title} {\enquote {\bibinfo {title} {Spin-glass properties of an
  {I}sing antiferromagnet on the {A}rchimedean $(3,12^2)$ lattice},}\ }\href
  {https://doi.org/10.1103/PhysRevB.72.024445} {\bibfield  {journal} {\bibinfo
  {journal} {Physical Review B}\ }\textbf {\bibinfo {volume} {72}},\ \bibinfo
  {pages} {024445} (\bibinfo {year} {2005})}\BibitemShut {NoStop}%
\bibitem [{\citenamefont {Krawczyk}\ \emph {et~al.}(2010)\citenamefont
  {Krawczyk}, \citenamefont {Malarz}, \citenamefont {Korff},\ and\
  \citenamefont {Ku{\l}akowski}}]{Krawczyk2010}%
  \BibitemOpen
  \bibfield  {author} {\bibinfo {author} {\bibfnamefont {M.~J.}\ \bibnamefont
  {Krawczyk}}, \bibinfo {author} {\bibfnamefont {K.}~\bibnamefont {Malarz}},
  \bibinfo {author} {\bibfnamefont {R.}~\bibnamefont {Korff}},\ and\ \bibinfo
  {author} {\bibfnamefont {K.}~\bibnamefont {Ku{\l}akowski}},\ }\bibfield
  {title} {\enquote {\bibinfo {title} {Communication and trust in the bounded
  confidence model},}\ }\href {https://doi.org/10.1007/978-3-642-16693-8_10}
  {\bibfield  {journal} {\bibinfo  {journal} {Lecture Notes in Computer
  Science}\ }\textbf {\bibinfo {volume} {6421}},\ \bibinfo {pages} {90--99}
  (\bibinfo {year} {2010})}\BibitemShut {NoStop}%
\bibitem [{\citenamefont {Malarz}, \citenamefont {Stauffer},\ and\
  \citenamefont {Ku{\l}akowski}(2006)}]{Stauffer2006}%
  \BibitemOpen
  \bibfield  {author} {\bibinfo {author} {\bibfnamefont {K.}~\bibnamefont
  {Malarz}}, \bibinfo {author} {\bibfnamefont {D.}~\bibnamefont {Stauffer}},\
  and\ \bibinfo {author} {\bibfnamefont {K.}~\bibnamefont {Ku{\l}akowski}},\
  }\bibfield  {title} {\enquote {\bibinfo {title} {Bonabeau model on a fully
  connected graph},}\ }\href {https://doi.org/10.1140/epjb/e2006-00059-3}
  {\bibfield  {journal} {\bibinfo  {journal} {European Physical Journal B}\
  }\textbf {\bibinfo {volume} {50}},\ \bibinfo {pages} {195--198} (\bibinfo
  {year} {2006})}\BibitemShut {NoStop}%
\bibitem [{\citenamefont {Franceschini}\ \emph {et~al.}(1999)\citenamefont
  {Franceschini}, \citenamefont {Hasinger}, \citenamefont {Miyaji},\ and\
  \citenamefont {Malquori}}]{example_4}%
  \BibitemOpen
  \bibfield  {author} {\bibinfo {author} {\bibfnamefont {A.}~\bibnamefont
  {Franceschini}}, \bibinfo {author} {\bibfnamefont {G.}~\bibnamefont
  {Hasinger}}, \bibinfo {author} {\bibfnamefont {T.}~\bibnamefont {Miyaji}},\
  and\ \bibinfo {author} {\bibfnamefont {D.}~\bibnamefont {Malquori}},\
  }\bibfield  {title} {\enquote {\bibinfo {title} {On the relationship between
  galaxy formation and quasar evolution},}\ }\href
  {https://doi.org/10.1046/j.1365-8711.1999.03078.x} {\bibfield  {journal}
  {\bibinfo  {journal} {Monthly Notices of the Royal Astronomical Society}\
  }\textbf {\bibinfo {volume} {310}},\ \bibinfo {pages} {L5--L9} (\bibinfo
  {year} {1999})}\BibitemShut {NoStop}%
\bibitem [{\citenamefont {Chinazzi}\ \emph {et~al.}(2019)\citenamefont
  {Chinazzi}, \citenamefont {Goncalves}, \citenamefont {Zhang},\ and\
  \citenamefont {Vespignani}}]{example_1}%
  \BibitemOpen
  \bibfield  {author} {\bibinfo {author} {\bibfnamefont {M.}~\bibnamefont
  {Chinazzi}}, \bibinfo {author} {\bibfnamefont {B.}~\bibnamefont {Goncalves}},
  \bibinfo {author} {\bibfnamefont {Q.}~\bibnamefont {Zhang}},\ and\ \bibinfo
  {author} {\bibfnamefont {A.}~\bibnamefont {Vespignani}},\ }\bibfield  {title}
  {\enquote {\bibinfo {title} {Mapping the physics research space: {A} machine
  learning approach},}\ }\href
  {https://doi.org/10.1140/epjds/s13688-019-0210-z} {\bibfield  {journal}
  {\bibinfo  {journal} {EPJ Data Science}\ }\textbf {\bibinfo {volume} {8}},\
  \bibinfo {pages} {33} (\bibinfo {year} {2019})}\BibitemShut {NoStop}%
\bibitem [{\citenamefont {Simini}\ \emph {et~al.}(2012)\citenamefont {Simini},
  \citenamefont {Gonz\'alez}, \citenamefont {Maritan},\ and\ \citenamefont
  {Barab\'asi}}]{example_3}%
  \BibitemOpen
  \bibfield  {author} {\bibinfo {author} {\bibfnamefont {F.}~\bibnamefont
  {Simini}}, \bibinfo {author} {\bibfnamefont {M.~C.}\ \bibnamefont
  {Gonz\'alez}}, \bibinfo {author} {\bibfnamefont {A.}~\bibnamefont
  {Maritan}},\ and\ \bibinfo {author} {\bibfnamefont {A.-L.}\ \bibnamefont
  {Barab\'asi}},\ }\bibfield  {title} {\enquote {\bibinfo {title} {A universal
  model for mobility and migration patterns},}\ }\href
  {https://doi.org/10.1038/nature10856} {\bibfield  {journal} {\bibinfo
  {journal} {Nature}\ }\textbf {\bibinfo {volume} {484}},\ \bibinfo {pages}
  {96--100} (\bibinfo {year} {2012})}\BibitemShut {NoStop}%
\bibitem [{\citenamefont {Goncalves}, \citenamefont {Balcan},\ and\
  \citenamefont {Vespignani}(2013)}]{example_2}%
  \BibitemOpen
  \bibfield  {author} {\bibinfo {author} {\bibfnamefont {B.}~\bibnamefont
  {Goncalves}}, \bibinfo {author} {\bibfnamefont {D.}~\bibnamefont {Balcan}},\
  and\ \bibinfo {author} {\bibfnamefont {A.}~\bibnamefont {Vespignani}},\
  }\bibfield  {title} {\enquote {\bibinfo {title} {Human mobility and the
  worldwide impact of intentional localized highly pathogenic virus release},}\
  }\href {https://doi.org/10.1038/srep00810} {\bibfield  {journal} {\bibinfo
  {journal} {Scientific Reports}\ }\textbf {\bibinfo {volume} {3}},\ \bibinfo
  {pages} {810} (\bibinfo {year} {2013})}\BibitemShut {NoStop}%
\bibitem [{\citenamefont {Eisler}, \citenamefont {Bartos},\ and\ \citenamefont
  {Kert\'esz}(2008)}]{example_5}%
  \BibitemOpen
  \bibfield  {author} {\bibinfo {author} {\bibfnamefont {Z.}~\bibnamefont
  {Eisler}}, \bibinfo {author} {\bibfnamefont {I.}~\bibnamefont {Bartos}},\
  and\ \bibinfo {author} {\bibfnamefont {J.}~\bibnamefont {Kert\'esz}},\
  }\bibfield  {title} {\enquote {\bibinfo {title} {Fluctuation scaling in
  complex systems: {T}aylor's law and beyond},}\ }\href
  {https://doi.org/10.1080/00018730801893043} {\bibfield  {journal} {\bibinfo
  {journal} {Advances in Physics}\ }\textbf {\bibinfo {volume} {57}},\ \bibinfo
  {pages} {89--142} (\bibinfo {year} {2008})}\BibitemShut {NoStop}%
\bibitem [{\citenamefont {Lima}\ and\ \citenamefont {Malarz}(2006)}]{Lima2006}%
  \BibitemOpen
  \bibfield  {author} {\bibinfo {author} {\bibfnamefont {F.~W.~S.}\
  \bibnamefont {Lima}}\ and\ \bibinfo {author} {\bibfnamefont {K.}~\bibnamefont
  {Malarz}},\ }\bibfield  {title} {\enquote {\bibinfo {title} {Majority-vote
  model on $(3,4,6,4)$ and $(3^4,6)$ {A}rchimedean lattices},}\ }\href
  {https://doi.org/10.1142/S0129183106009849} {\bibfield  {journal} {\bibinfo
  {journal} {International Journal of Modern Physics C}\ }\textbf {\bibinfo
  {volume} {17}},\ \bibinfo {pages} {1273--1283} (\bibinfo {year}
  {2006})}\BibitemShut {NoStop}%
\bibitem [{dat(2023)}]{data-intercont-rank}%
  \BibitemOpen
  \href {http://www.zis.agh.edu.pl/collaboration/Prank_continents.csv}
  {\enquote {\bibinfo {title}
  {http://www.zis.agh.edu.pl/\-collaboration/{P}rank\-\_continents.csv},}\ }
  (\bibinfo {year} {2023})\BibitemShut {NoStop}%
\bibitem [{\citenamefont {Newman}(2005)}]{zipf_mejn}%
  \BibitemOpen
  \bibfield  {author} {\bibinfo {author} {\bibfnamefont {M.~E.~J.}\
  \bibnamefont {Newman}},\ }\bibfield  {title} {\enquote {\bibinfo {title}
  {Power laws, {P}areto distributions and {Z}ipf's law},}\ }\href
  {https://doi.org/10.1080/00107510500052444} {\bibfield  {journal} {\bibinfo
  {journal} {Contemporary Physics}\ }\textbf {\bibinfo {volume} {46}},\
  \bibinfo {pages} {323--351} (\bibinfo {year} {2005})}\BibitemShut {NoStop}%
\bibitem [{\citenamefont {Lü}, \citenamefont {Zhang},\ and\ \citenamefont
  {Zhou}(2010)}]{heap_lu}%
  \BibitemOpen
  \bibfield  {author} {\bibinfo {author} {\bibfnamefont {L.}~\bibnamefont
  {Lü}}, \bibinfo {author} {\bibfnamefont {Z.-K.}\ \bibnamefont {Zhang}},\
  and\ \bibinfo {author} {\bibfnamefont {T.}~\bibnamefont {Zhou}},\ }\bibfield
  {title} {\enquote {\bibinfo {title} {Zipf's law leads to {H}eaps' law:
  {A}nalyzing their relation in finite-size systems},}\ }\href
  {https://doi.org/10.1371/journal.pone.0014139} {\bibfield  {journal}
  {\bibinfo  {journal} {PLoS One}\ }\textbf {\bibinfo {volume} {5}},\ \bibinfo
  {pages} {e14139} (\bibinfo {year} {2010})}\BibitemShut {NoStop}%
\bibitem [{\citenamefont {Font-Clos}, \citenamefont {Boleda},\ and\
  \citenamefont {Álvaro Corral}(2013)}]{heap_fc}%
  \BibitemOpen
  \bibfield  {author} {\bibinfo {author} {\bibfnamefont {F.}~\bibnamefont
  {Font-Clos}}, \bibinfo {author} {\bibfnamefont {G.}~\bibnamefont {Boleda}},\
  and\ \bibinfo {author} {\bibnamefont {Álvaro Corral}},\ }\bibfield  {title}
  {\enquote {\bibinfo {title} {A scaling law beyond {Z}ipf's law and its
  relation to {H}eaps' law},}\ }\href
  {https://doi.org/10.1088/1367-2630/15/9/093033} {\bibfield  {journal}
  {\bibinfo  {journal} {New Journal of Physics}\ }\textbf {\bibinfo {volume}
  {15}},\ \bibinfo {pages} {093033} (\bibinfo {year} {2013})}\BibitemShut
  {NoStop}%
\bibitem [{\citenamefont {Cameron}(2022)}]{cameron2022zipf}%
  \BibitemOpen
  \bibfield  {author} {\bibinfo {author} {\bibfnamefont {M.~P.}\ \bibnamefont
  {Cameron}},\ }\href {https://ideas.repec.org/p/wai/econwp/22-07.html}
  {\enquote {\bibinfo {title} {{Zipf's law across social media}},}\ }\bibinfo
  {type} {Working Papers in Economics}\ \bibinfo {number} {22/07}\ (\bibinfo
  {institution} {University of Waikato},\ \bibinfo {year} {2022})\BibitemShut
  {NoStop}%
\bibitem [{\citenamefont {Perotti}\ and\ \citenamefont
  {Billoni}(2020)}]{PEROTTI2020124309}%
  \BibitemOpen
  \bibfield  {author} {\bibinfo {author} {\bibfnamefont {J.~I.}\ \bibnamefont
  {Perotti}}\ and\ \bibinfo {author} {\bibfnamefont {O.~V.}\ \bibnamefont
  {Billoni}},\ }\bibfield  {title} {\enquote {\bibinfo {title} {On the
  emergence of {Z}ipf’s law in music},}\ }\href
  {https://doi.org/10.1016/j.physa.2020.124309} {\bibfield  {journal} {\bibinfo
   {journal} {Physica A: Statistical Mechanics and its Applications}\ }\textbf
  {\bibinfo {volume} {549}},\ \bibinfo {pages} {124309} (\bibinfo {year}
  {2020})}\BibitemShut {NoStop}%
\bibitem [{\citenamefont {Pietronero}\ \emph {et~al.}(2001)\citenamefont
  {Pietronero}, \citenamefont {Tosatti}, \citenamefont {Tosatti},\ and\
  \citenamefont {Vespignani}}]{zipf_vesp}%
  \BibitemOpen
  \bibfield  {author} {\bibinfo {author} {\bibfnamefont {L.}~\bibnamefont
  {Pietronero}}, \bibinfo {author} {\bibfnamefont {E.}~\bibnamefont {Tosatti}},
  \bibinfo {author} {\bibfnamefont {V.}~\bibnamefont {Tosatti}},\ and\ \bibinfo
  {author} {\bibfnamefont {A.}~\bibnamefont {Vespignani}},\ }\bibfield  {title}
  {\enquote {\bibinfo {title} {Explaining the uneven distribution of numbers in
  nature: {T}he laws of {B}enford and {Z}ipf},}\ }\href
  {https://doi.org/10.1016/S0378-4371(00)00633-6} {\bibfield  {journal}
  {\bibinfo  {journal} {Physica A}\ }\textbf {\bibinfo {volume} {293}},\
  \bibinfo {pages} {297--304} (\bibinfo {year} {2001})}\BibitemShut {NoStop}%
\bibitem [{\citenamefont {James}\ \emph {et~al.}(2018)\citenamefont {James},
  \citenamefont {Azaele}, \citenamefont {Maritan},\ and\ \citenamefont
  {Simini}}]{PhysRevE.98.032408}%
  \BibitemOpen
  \bibfield  {author} {\bibinfo {author} {\bibfnamefont {C.}~\bibnamefont
  {James}}, \bibinfo {author} {\bibfnamefont {S.}~\bibnamefont {Azaele}},
  \bibinfo {author} {\bibfnamefont {A.}~\bibnamefont {Maritan}},\ and\ \bibinfo
  {author} {\bibfnamefont {F.}~\bibnamefont {Simini}},\ }\bibfield  {title}
  {\enquote {\bibinfo {title} {{Z}ipf's and {T}aylor's laws},}\ }\href
  {https://doi.org/10.1103/PhysRevE.98.032408} {\bibfield  {journal} {\bibinfo
  {journal} {Physical Review E}\ }\textbf {\bibinfo {volume} {98}},\ \bibinfo
  {pages} {032408} (\bibinfo {year} {2018})}\BibitemShut {NoStop}%
\bibitem [{\citenamefont {Blasius}\ and\ \citenamefont
  {T\"onjes}(2009)}]{PhysRevLett.103.218701}%
  \BibitemOpen
  \bibfield  {author} {\bibinfo {author} {\bibfnamefont {B.}~\bibnamefont
  {Blasius}}\ and\ \bibinfo {author} {\bibfnamefont {R.}~\bibnamefont
  {T\"onjes}},\ }\bibfield  {title} {\enquote {\bibinfo {title} {Zipf's law in
  the popularity distribution of chess openings},}\ }\href
  {https://doi.org/10.1103/PhysRevLett.103.218701} {\bibfield  {journal}
  {\bibinfo  {journal} {Physical Review Letters}\ }\textbf {\bibinfo {volume}
  {103}},\ \bibinfo {pages} {218701} (\bibinfo {year} {2009})}\BibitemShut
  {NoStop}%
\end{thebibliography}%

\end{document}